# COMPARING THE PREDICTED ABELL/ACO CLUSTER AND THE MKIII GALAXY DENSITY & VELOCITY FIELDS

M. PLIONIS
*National Observatory of Athens, Thesio, 18110 Athens, Greece*

E. BRANCHINI
*Department of Physics, University of Durham, DH1 3LE, UK*

AND

I. ZEHAVI AND A. DEKEL
*The Racah Institute of Physics, The Hebrew University, Givat Ram, Jerusalem 91904, Israel*

**Abstract.** We compare the predicted Abell/ACO cluster density & velocity fields, using linear theory, with that based on the MkIII catalog and the POTENT machinery within $R \sim 7000$ km/sec. Both the density & velocity fields are in good qualitative agreement indicating that apart from a linear biasing factor the Abell/ACO clusters do trace the underlying mass distribution quite well.

## 1. Introduction

In two recent papers (Branchini & Plionis 1996 and Branchini, Plionis & Sciama 1996) it has been shown that using linear theory, linear biasing and a dynamical algorithm it was possible to reconstruct the real-space Abell/ACO cluster density field and thus their peculiar velocity field. Among other things it was found that: **(a)** the LG motion is reflected in the cluster inertial frame, in disagreement with Lauer & Postman 1994 and **(b)** the cluster bulk motion is consistent on the relevant scales with that derived from galaxy peculiar velocity catalogs (c.f. Dekel 1994; Da Costa et al. 1996).

Here we present a further and direct test of the validity of the hypothesis that Abell/ACO clusters of galaxies are reliable tracers of the underlying mass, by comparing qualitatively their reconstructed density and velocity fields with the corresponding underlying fields as obtained by the POTENT method (Dekel, Bertschinger & Faber 1990, Bertschinger et al. 1990) applied to the Mark III catalog of peculiar velocities (cf. Willick et al 1995).

## 2. Comparing Clusters & Mark III - POTENT velocity and density fields

For the purpose of the comparison, the reconstructed cluster and MKIII density and velocity fields have been computed onto a 320 $h^{-1}$ Mpc cubic grid with 5 $h^{-1}$ Mpc spacing centered onto the LG location. However, the comparison will be performed only within a sphere of 70 $h^{-1}$ Mpc around the LG, where the MKIII catalog is most reliable. Numerical tests performed suggest that a Gaussian window with a 15 $h^{-1}$ Mpc radius is the optimal window for comparing the cluster and the POTENT fields ($c$ and $P$ fields hereafter). The details of the smoothing procedure and of the simulations performed to test the robustness of our results will be presented in Plionis et al. (1997).

Figure 1 displays the smoothed $c$ and $P$ density fluctuation fields (see caption for details). The similarity between the two fields is evident, especially that of the supergalactic plane (middle row) in which the Great Attractor region and the Perseus-Pisces supercluster can be recognized in both the $P$ and $c$ fields as density peaks on the left and right side of the plots, respectively. These two structures appear, in both cases, separated by an extended underdense region. However, the Coma supercluster seen in the $c$-field at $(X,Y)_{sup} \approx (0,60)$ $h^{-1}$ Mpc, is missed in the $P$-field with some evidence for a positional displacement. Similar qualitative agreement appears also in the upper slice while discrepancies between the $P$ and $c$ density fields are evident in the lower slice at positive supergalactic $X$s. On the other hand the comparison of the $P$ and $c$ velocity fields show a less remarkable agreement (Fig.2). The main feature common to the two velocity fields is the infall onto the GA region, while the infall onto the Perseus Pisces overdensity, predicted from the cluster distribution, is substituted by a general motion of this region towards the LG in the $P$-field. The divergent $P$-flow on the top right corner is much less significant in the $c$-field where, in general, the velocities' amplitudes are smaller than the $P$-ones. Another feature absent in the $P$- field is the infall to the Coma region apparent in the $c$-field. It should be noted though that the largest discrepancies lie near the volume borders where the number of Mark III objects is small and their distance estimation less reliable.

The velocity-velocity (**v**-**v**) comparison, being non-local, allows one to account for the mass distribution up to very large scales and thus use all the cluster information, but its reliability is reduced by the ill-sampled regions (e.g. the zone of avoidance); while the density-density ($\delta$-$\delta$) comparison, being local, is less sensitive to this, but suffers from the paucity of clusters within the regions in which the fields are compared. The derived value of the $\beta_c (\equiv \Omega_o^{0.6}/b_c)$ from both these comparisons and all relevant tests will be presented in Plionis et al. (1997).

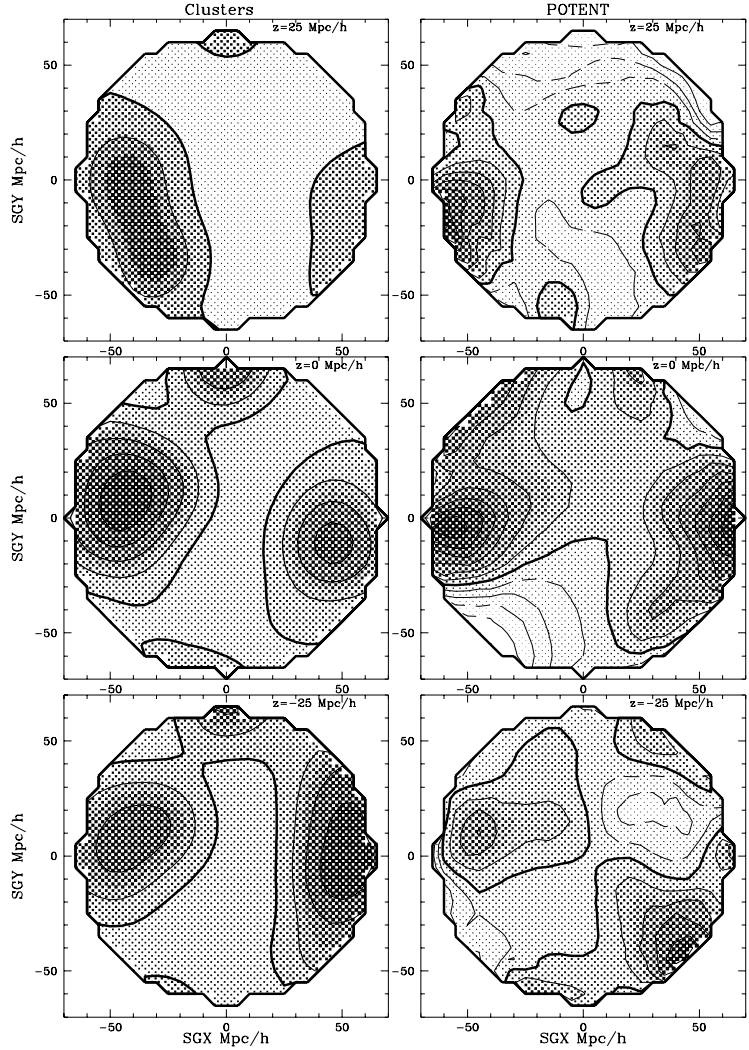

*Figure 1.* The overdensity field in slices parallel to the supergalactic plane, smoothed with a Gaussian filter of radius 1500 km/sec. The plots in the left hand column refers to the cluster field while the MkIII-POTENT mass field is shown in the right hand plots. The contour levels are in step of $\Delta\delta = 0.2$, solid contours refer to overdense regions while dashed contours refer to negative overdensities. The thick line indicates the $\delta = 0$ contour. The supergalactic plane slice is displayed in the middle row while the upper and lower rows refer to the supergalactic $Z = 2500$ km/sec and $Z = -2500$ km/sec, respectively. The amplitude of the c-field has been divided by $b_c = 4.8$.

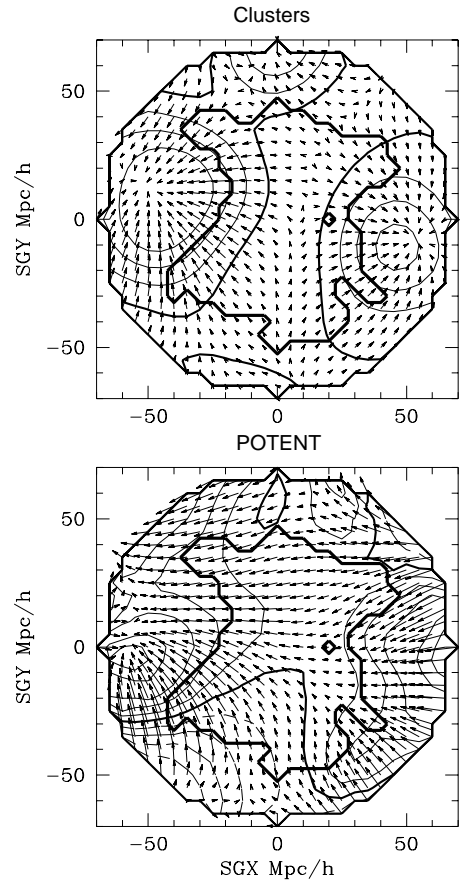

*Figure 2.* The supergalactic plane $c$ and $P$ velocity fields (in the CMB frame), superimposed to their overdensity contours. The heavy thick line represents the boundary of the 'reliable' volume. Note that the displayed $c$–field velocities are normalized to $\beta_c = 0.21$.

## References

Bertschinger E., Dekel A., Faber, S.M., Dressler A., Burstein D., 1990, ApJ, 364, 370
Branchini E., Plionis M., 1996, ApJ, 460, 569
Branchini E., Plionis M., Sciama D.W., 1996, ApJ, 461, L17
Da Costa et al. 1996, in eds. Balkowski C., Maurogordato S., Tao C., Trân Thanh Vân, Proc. of the Moriond Meeting on Clustering in the Universe (in press)
Dekel A., 1994, ARA&A, 32, 371
Dekel A., Bertschinger E., Faber S.M., 1990, ApJ, 364, 349
Lauer, T.R. & Postman, M., 1994, ApJ, 425, 418
Plionis, M., Branchini, E., Zehavi, I. & Dekel, A., 1997, in preparation
Willick, J.A., Courteau, S., Faber, S.M., Burstein, D. & Dekel, A., 1995, ApJ,446, 12